\documentclass[preprint]{aastex63}
\received{June 1, 2019}
\revised{January 10, 2019}
\accepted{\today}
\submitjournal{AJ}

\shorttitle{Tidal fragmentation of a star into planets 
around a supermassive black hole}
\shortauthors{Hayasaki et al.}
\usepackage{color}

\def\vec#1{\mbox{\boldmath $#1$}}

\begin{document}

\title{Ionization and dissociation induced fragmentation of a tidally disrupted star 
into planets around a supermassive black hole\\
}

\author[0000-0003-4799-1895]{Kimitake Hayasaki}
\affiliation{Department of Astronomy and Space Science, Chungbuk National University, Cheongju 361-763, Republic of Korea}

\author{Matthew R. Bate}
\affiliation{School of Physics, University of Exeter, Stocker Road, Exeter EX4 4QL}


\author{Abraham Loeb}
\affiliation{Harvard-Smithsonian Center for Astrophysics, 60 GardenStreet, Cambridge, MA02138, USA}







\begin{abstract}
We show results from the radiation hydrodynamics (RHD) simulations of tidal 
disruption of a star on a parabolic orbit by a supermassive black hole (SMBH) 
based on a three-dimensional smoothed particle hydrodynamics code with 
radiative transfer. We find that such a tidally disrupted star fragment and form 
clumps soon after its tidal disruption. The fragmentation results from the 
endothermic processes of ionization and dissociation that reduce the gas pressure, 
leading to local gravitational collapse. Radiative cooling is less effective because 
the stellar debris is still highly optically thick in such an early time.
Our simulations reveal that a solar-type star with a stellar density profile 
of $n=3$ disrupted by a $10^6$ solar mass black hole produces $\sim20$ 
clumps of masses in the range of $0.1$ to $12$ Jupiter masses.
The mass fallback rate decays with time, with pronounced spikes 
from early to late time. The spikes provide evidence for the clumps of the 
returning debris, while the clumps on the unbound debris can be potentially 
freely-floating planets and brown dwarfs. This ionization and dissociation induced 
fragmentation on a tidally disrupted star are a promising candidate mechanism to 
form low-mass stars to planets around an SMBH.
\end{abstract}



\section{Introduction} 

%
%
Supermassive black holes (SMBHs) reside ubiquitously at the center of 
galaxies, based on observations of stellar proper motion, stellar velocity 
dispersion or accretion luminosity \citep{2013ARAA..51..511K}. 
Tidal disruption events (TDEs) are thought to be a key signature of dormant SMBHs 
at the centers of the inactive galaxies or intermediate-mass black holes (IMBHs) in star clusters. 
As a star approaches to a SMBH, it is torn apart by the tidal force of the black hole, 
which dominates the self-gravity of the star at the tidal disruption radius:
\begin{equation}
 r_{\rm t}=\left(\frac{M_{\rm bh}}{m_*}\right)^{1/3}r_{*}\approx
24\left(\frac{M_{\rm bh}}{10^6\,{M}_\odot}\right)^{-2/3}
\left(\frac{m_*}{{M}_\odot}\right)^{-1/3}
\left(\frac{r_*}{{R}_\odot}\right)
r_{\rm S}.
\label{eq:rt}
\end{equation} 
Here we denote the black hole mass with $M_{\rm bh}$, stellar mass 
with $m_*$ and radius with $r_*$, and the Schwarzschild radius 
with $r_{\rm{S}}=2{\rm G}M_{\rm bh}/{\rm c}^2$, where $G$ and 
$c$ are Newton's gravitational constant and the speed of light, 
respectively. 
Most TDEs take place when a star at large separation ($\sim1\,\rm{pc}$) is perturbed onto 
a parabolic orbit approaching close enough to the SMBH to be ripped apart by its tidal force. 
The subsequent accretion of stellar debris falling back to the SMBH produces a characteristic 
flare with a luminosity that could exceed the Eddington luminosity for a time scale of weeks 
to months \citep{rees88,p89,ek89}.

%
%
TDE flares have been discovered at optical, ultraviolet, and soft X-ray 
(see \citealt{2015JHEAp...7..148K} for a review; \citealt{2016MNRAS.455.2918H}; 
\citealt{2017ApJ...838..149A}) wavelengths with infrared event rates of $10^{-5}-10^{-4}$ 
per year per Milky-way mass galaxy \citep{dbeb02,sm16}, while jetted TDEs have been 
detected through non-thermal emissions in radio \citep{baz+11,2016ApJ...819L..25A,2016Sci...351...62V} 
or hard X-ray \citep{dnb+11,2015MNRAS.452.4297B}, with a lower event rate 
\citep{2014arXiv1411.0704F,2017MNRAS.469.1354D}. Some spectroscopic 
researches confirmed ${\rm H \,I}$ and ${\rm He\, II}$ (\citealt{ar14}) as well as 
metal lines (\citealt{2019arXiv190303120L}) in TDEs. Weakly relativistic blue-shifted 
broad absorption lines were attributed to a high-velocity outflow produced in TDE 
AT2018zr (\citealt{2019arXiv190305637H}).

%
%
It is still debated how the expected fallback rate as a function of time, 
$\propto{t}^{-5/3}$, translates into the observed light curves. While most of 
the soft X-ray TDEs appear to follow the $t^{-5/3}$ scaling, the optical to 
ultraviolet TDEs exhibit the different decay (e.g., \citealt{2012Natur.485..217G}).
\cite{lkp09} numerically showed that the fallback rate depends on the internal structure 
of the tidally disrupted star, leading to early-time deviations from the standard behavior.
The centrally condensed core survived by the partial disruption of the star can make 
the light curve steeper \citep{2013ApJ...767...25G}. 
The accretion of clumps formed by self-gravitational fragmentation of the debris stream 
causes the significant variations of the light curve around the $t^{-5/3}$ average at very 
late time \citep{2015ApJ...808L..11C}, although they did not explain the underlying 
trigger for the self-gravitating instability. 
These clumps could be the origin of G2 cloud observed around Sgr A$^*$ 
\citet{2014ApJ...786L..12G},
which made the clumpy structure on the debris by a fluid instabilities caused by the interaction between the debris stream and an ambient medium.

%
%
In this paper, we explore the possibility of rapid fragmentation of a tidally disrupted star 
around a SMBH through radiation hydrodynamic (RHD) simulations. 
In section~\ref{sec:2}, we describe our numerical approach, focusing on the radiation transfer we make use of. In section~\ref{sec:3}, we study what causes rapid fragmentation of stellar debris, and predict 
the resulting fallback rate.
Finally, section~\ref{sec:con} concludes with a discussion of our results and their implication.

%
\section{Computational Method}
\label{sec:2}
%

%
%
We start by describing our numerical methods, with a focus on how 
we handle the radiation transfer in the numerical code, and summarize the setup of our physical 
and numerical models. The simulations presented below were performed using a 
three-dimensional (3D), Smoothed Particle Hydrodynamics (SPH) code with radiative transfer.
The code is based on the original version of \cite{benz90a,benz90b}, but substantially 
modified as described in \cite{bate95} and parallelized using both OpenMP and MPI.
Subsequently, the radiation hydrodynamics was incorporated into the code by \cite{wb04} 
and \cite{wbm05}, and it has been used extensively to study star formation (e.g \citealt{wb06,2012MNRAS.419.3115B}).

%
%


%
\subsection{Equations of radiation hydrodynamics in SPH} 
\label{sec:radSPH}
%
In a frame comoving with a radiating fluid, assuming the local thermodynamics equilibrium (LTE), 
the equations of self-gravitating, non-viscous RHD to order unity in $v/c$ are given by 
\citep{ts01,wb04,wbm05},
\begin{eqnarray}
&&
\frac{D\rho}{Dt}=-\rho\vec\nabla\cdot\vec{v}, 
\label{eq:masscons}
\\
&&
\rho\frac{D\vec{v}}{dt}=-\vec\nabla{p}+\frac{\chi_{\rm F}\rho}{c}\vec{F}-\rho\vec\nabla\phi, 
\label{eq:momcons}
\\
&&
\rho\frac{D}{Dt}\left(\frac{e}{\rho}\right)=-p\vec\nabla\cdot\vec{v}-4\pi\kappa_{\rm p}\rho{B}+c\kappa_{\rm E}\rho{E},
\label{eq:gasene}
\\
&&
\rho\frac{D}{Dt}\left(\frac{E}{\rho}\right)=-\vec{\nabla}\cdot{\vec{F}}-\vec\nabla\vec{v}:\vec{P}+4\pi\kappa_{\rm p}\rho{B}-c\kappa_{\rm E}\rho{E},
\label{eq:radene}
\\
&&
\frac{\rho}{c^{2}}\frac{D}{Dt}\left(\frac{\vec{F}}{\rho}\right)=-\nabla\cdot\vec{P}-\frac{\chi_{\rm F}\rho}{c}\vec{F}, 
\label{eq:radflux}
\\
&&
\nabla^{2}\phi=4\pi{G}\rho,
\label{eq:selfgrav}
\end{eqnarray}
where $D/Dt\equiv\partial/\partial{t}+\vec{v}\cdot\vec\nabla$ is the convective derivative, 
$\rho$ is the density, $\vec{v}$ is the velocity, $u$ is the specific energy of the gas, 
$p$ is the scalar isotropic pressure of the gas, $\phi$ is the gravitational potential, 
$B$ is the frequency-integrated Planck function, $E$ is the frequency-integrated radiation energy, 
$\vec{F}$ is the momentum flux density, $\vec{P}$ is the radiation pressure tensor.
The colon product: \vec\nabla\vec{v}:\vec{P} indicates contraction over two indices as follows:
$(\partial{v}_{j}/\partial{x}_i)P_{ij}$.

Note that equations (3)-(5) have been integrated over frequency. This leads to the 
flux mean opacity $\chi_{\rm F}$, the Planck mean opacity $\kappa_{\rm P}$, 
and the energy mean opacity $\kappa_{\rm E}$. The opacities are assumed to be 
independent of frequency so that $\kappa_{\rm P}=\kappa_{\rm E}$, and they are newly 
defined as $\kappa$ without the subscripts. The total opacity, $\chi_{\rm F}$, should be the 
sum of the components of $\kappa$ and scattering. In our simulations, we ignore the 
scattering so that $\chi_{\rm F}=\kappa$.

%
\subsubsection{Flux-limited diffusion}
%

The flux-limited diffusion method provides the following relation \citep{lp81},
\begin{equation}
\vec{F}=-\mathcal{D}\nabla{E},
\label{eq:fluxdiffusion}
\end{equation}
with a diffusion constant $\mathcal{D}=c\lambda(E)/(\chi\rho)$, where $\lambda(E)$ 
is the flux limiter, as an alternative of equation~(\ref{eq:radflux}). 
It assumes a radiation pressure stress tensor,
\begin{equation}
\vec{P}=\vec{f}E,
\end{equation}
where $\vec{f}$ is the Eddington tensor:
\begin{equation}
\vec{f}=\frac{1}{2}(1-f(E))\vec{I}+\frac{1}{2}(3f(E)-1)\vec{\hat{n}}\vec{\hat{n}},
\end{equation}
with isotropic unit tensor $\vec{I}$ and unit vector $\vec{\hat{n}}=\vec\nabla{E}/|\nabla\vec{E}|$ 
in the direction of radiation energy gradient. Here, $f(E)$ is the Eddington factor: 
\begin{equation}
f(E)=\lambda(E)+\lambda^2(E)\mathcal{R}(E)^2,
\end{equation}
where $\mathcal{R}(E)=|\nabla{E}|/(\chi\rho{E})$ is the dimensionless quantity and 
we choose the flux limiter of \cite{lp81}:
\begin{equation}
\lambda(E)=\frac{2+\mathcal{R}(E)}{6+3\mathcal{R}(E)+\mathcal{R}^2(E)}.
\label{eq:fluxlimiter}
\end{equation} 
We solve the RHD equations by using equations (\ref{eq:fluxdiffusion})-(\ref{eq:fluxlimiter}) 
without solving equation (\ref{eq:radflux}) directly.

%
\subsubsection{Equation of state and opacities}
%
In order to close the equations (\ref{eq:masscons})-(\ref{eq:fluxlimiter}), 
we need to add the equation of state for the gas:
\begin{equation}
p=\mathcal{E}(T)\rho
\label{eq:eqs}
\end{equation}
where $\mathcal{E}(T)$ is a specific internal energy. 
Note that $\mathcal{E}(T)$ equals to $(R_{\rm g}/\mu)T$ for an ideal gas, 
where $R_{\rm g}$ is the gas constant and $\mu$ is the molecular weight.
For pure hydrodynamic simulations an adiabatic equation of state 
\begin{equation}
p=K\rho^{\gamma}
\end{equation}
is used in addition to the ideal equation of state, 
where $\gamma$ is the specific heat ratio ($\gamma=5/3$ is applied for a monoatomic gas) 
and $K$ is a proportionality constant. On the other hand, when the ionization and dissociation of the molecular hydrogen and hydrogen is included, \cite{bb75} derived:
\begin{eqnarray}
\mathcal{E}(T)
&=&
X(1-y)E(H_2)T_{\rm}+[1.5X(1+x)y+0.375Y(1+z_1+z_1z_2)]R_{\rm g}T_{\rm}
\nonumber \\
&+&
X(1.304\times10^{13}x+2.143\times10^{12})y
+Y(5.888\times10^{12}(1-z_2)+1.892\times10^{13}z_2)z_1,
\label{eq:cv2}
\end{eqnarray}
where X and Y are the mass fractions of hydrogen and helium, respectively, 
$y$ is the dissociation fraction of hydrogen, x the ionization fraction of hydrogen, 
and $z_1$ and $z_2$ are the degrees of single and double ionization of helium, respectively. 
$E(H2)$ gives the contribution to the specific heat capacity from molecular hydrogen and the 
ionization fractions are calculated using the standard Saha equation \citep{bb75,2007ApJ...656L..89B}. 
The mean molecular weight also changes with the degree of the 
ionization and dissociation, and was derived by \cite{bb75}, 
\begin{eqnarray}
\mu=\frac{4}{2X(1+y+2xy)+Y(1+z_1+z_2)}.
\label{eq:mu}
\end{eqnarray}
In our simulations, the solar elemental composition (X = 0.70 and Y = 0.28) are applied and the mean molecular weight of the gas is initially $\mu=2.38$ for this composition \citep{2012MNRAS.419.3115B}.

We adopt the gas opacity tables from \cite{alex75} and the dust opacity from \cite{pmc85}, and 
from \cite{fer05} at the higher temperatures when the dust is sublimated. Above $10^4$ K, 
the opacity is dominated by the Kramers law and electron scattering: 
\begin{equation}
\kappa=\kappa_{\rm K}+\kappa_{\rm es}=1.2512\times10^{22}\rho{T}^{-7/2}+0.4\,{\rm cm^2\,g^{-1}}.
\end{equation}
For temperatures above a few million K, the opacity, specific heat capacity, and mean molecular mass approach constant values. 

%
\subsection{Modeling stellar tidal disruption 
by SPH with radiative transfer}
\label{sec:models}
%

We follow two-stages in studying the process 
of tidal disruption of a vilialized star by a SMBH.
In the first stage, we model a star by the polytrope, which is a solution to the 
Lane-Emden equation with a polytropic index $n=1.5$ and $n=3$ as an initial condition.
We then run five types of simulations of a solar-type star with mass and radius 
$(m_*,r_*)=(1\,M_\odot,1\,R_\odot)$, and a solar metallicity ($1\,Z_\odot$) 
for the RHD cases.
We finish the simulations at $t=50$ when they are well-virialized, where the unit of time 
is given by $\Omega_{*}^{-1}=\sqrt{r_*^3/Gm_*}\simeq1.6\times10^3\,\rm{s}$.
Models S1 and S3 show the purely hydrodynamic simulations with $n=1.5$ and $n=3$, 
respectively, whereas Model S2 represents the RHD simulation of the polytrope of $n=1.5$. 
Models S4 and S5 do the RHD simulations of an $n=3$ polytrope with and without radiative transfer.
The details of each model are shown in Table~\ref{tbl:tdes}.
For all models, the magnitude ratio of thermal to gravitational 
energies ranges between $0.497 \le |E_{\rm th}/E_{\rm grav}|\le0.499$ 
at $t=0$ and takes a value of $|E_{\rm th}/E_{\rm grav}|=0.497$ at $t=50$.
It means that stars that we make use of in the second-stage are well-virialized.

\begin{figure}
\resizebox{\hsize}{!}{
\includegraphics*[width=5cm]{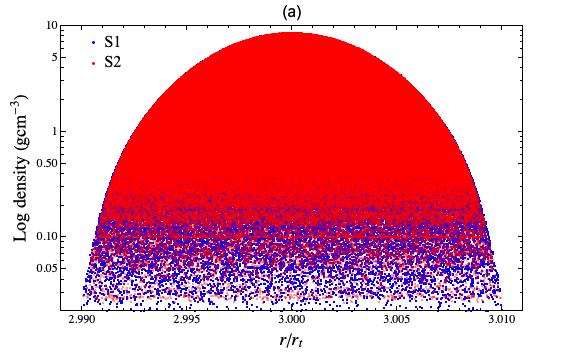}
\includegraphics*[width=5cm]{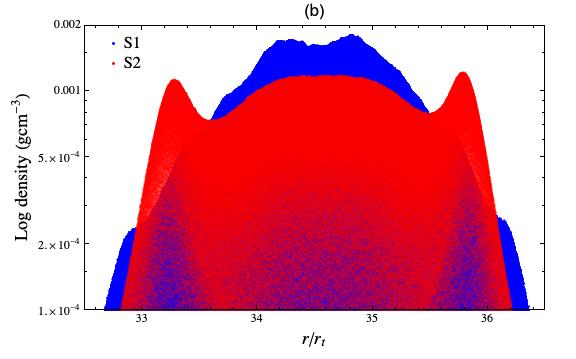}
}\\
\resizebox{\hsize}{!}{
\includegraphics*[width=5cm]{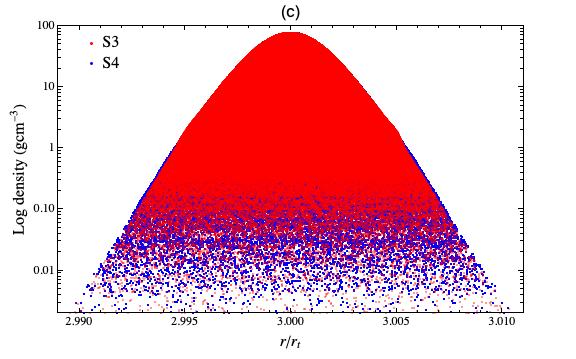}
\includegraphics*[width=5cm]{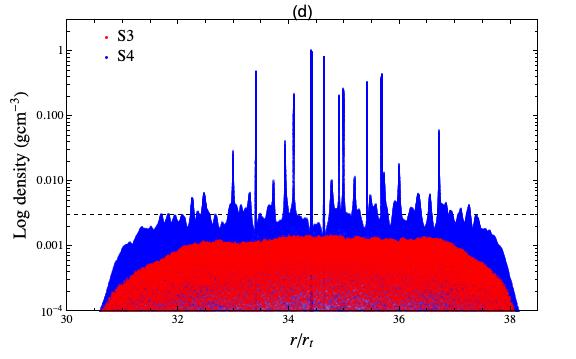}
}\\
\resizebox{\hsize}{!}{
\includegraphics*[width=5cm]{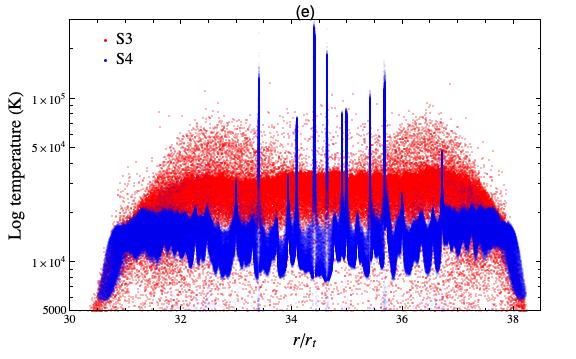}
\includegraphics*[width=5cm]{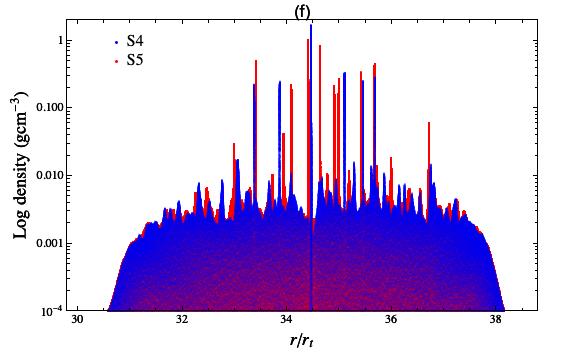}
}
\caption{
Radial density distributions of the star at $t=0$ and of the stellar debris at $t=100$. 
Panel (a) shows the overlapped density distribution between Models S1 and S2 at t=0, 
while panel (b) shows the same distribution as panel (a) but for at t=100. 
Panel (c) shows the overlapped density distribution between Models S3 and S4 at t=0, 
while panel (d) shows the same distribution as panel (c) but for at t=100.
In panel (d), the dashed line represents $\rho_{\rm crit}=3\times10^{-3}\,{\rm g\,cm^{-3}}$, 
which is set as a fiducial value of the gas density. We then recognize what has the density beyond 
it as a clump. Panel (e) depicts the corresponding temperature distribution to panel (d).
Panel (f) represents the overlapped density distribution between Models S4 and S5 at $t=100$.
}
\label{fig:raddis}
\end{figure}

In the second stage, the SMBH is added at the origin and the star is initially 
located at the distance of three times its tidal disruption radius from the black hole. 
The total number of SPH particles used in each simulation is $\sim10^6$. 
In all second-stage simulations, we used $M_{\rm bh}=10^6 M_{\odot}$, $m_*=1\,M_\odot$, 
$r_*=1\,R_\odot$, and termination time of each simulation is $t=100$. Figure~\ref{fig:raddis} depicts the radial density distribution of the stellar debris for the five models and the radial temperature distribution of Models 3 and 4. Panel (a) represents that of Models S1 and S2 at $t=0$, whereas panel (b) does that of Models S1 and S2 at $t=100$. Panel (c) represents that of Models S3 and S4 at $t=0$, whereas panel (d) does that of Models S3 and S4 at $t=100$. Panel (e) represents the corresponding temperature distribution to panel (d). For comparison, panel (d) represnets the overlapped density distribution between those of Models S4 and S5. Panel (a) shows that Model S1 initially have the same density profile as Model S2, while panel (c) shows that Models S3 initially have the same density profile as Model S4. Note that Model S5 has also the same density profile at $t=0$ as those of Models S3 and S4.

%
%
\begin{table}
\centering
\caption{
Simulated models of TDEs. 
The first column shows the model label. 
The second column is the polytrope index, $n$, of gas sphere 
at the initial setting. The third column indicates whether radiative 
transfer was used. The fourth and fifth columns provide the number 
of clumps and the clump's mass, respectively. The last column 
describes the type of corresponding simulation.
}
  \begin{tabular}{@{}ccccccccccl@{}}
  \hline
Model 
& Polytropic index
& Radiative transfer
& Fragmentation
& Clumpy number
& Remark \\
& $n$ &  & at $t=100$ 
& $N_{\rm cl}$   
\\
\hline
S1 & $1.5$ &  $-$ & no & $0$ & Pure Hydro.\\
S2 & $1.5$ & on   & no & $0$ & RHD  \\
S3 & $3$    &  $-$ & no & $0$ & Pure Hydro.\\
S4 & $3$    & on   & yes & $16$ & RHD \\
S5 & $3$    & off   & yes & $19$ & RHD \\
\hline
\end{tabular}
\label{tbl:tdes}
\end{table}

%
\section{Fragmentation of the stellar debris}
\label{sec:3}
%

\begin{figure}
\resizebox{\hsize}{!}{\includegraphics*[width=9cm]{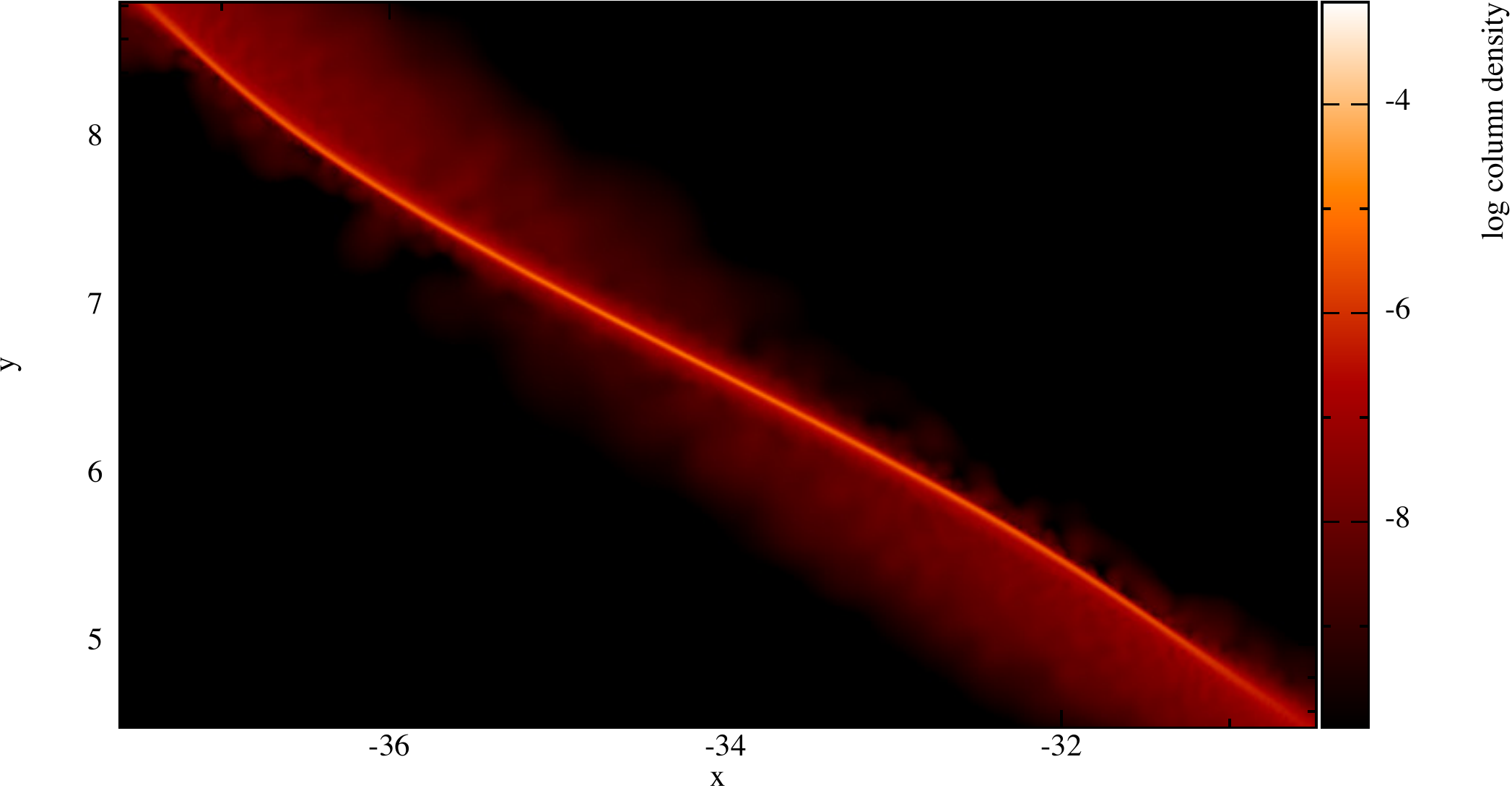}}\\
\resizebox{\hsize}{!}{\includegraphics*[width=9cm]{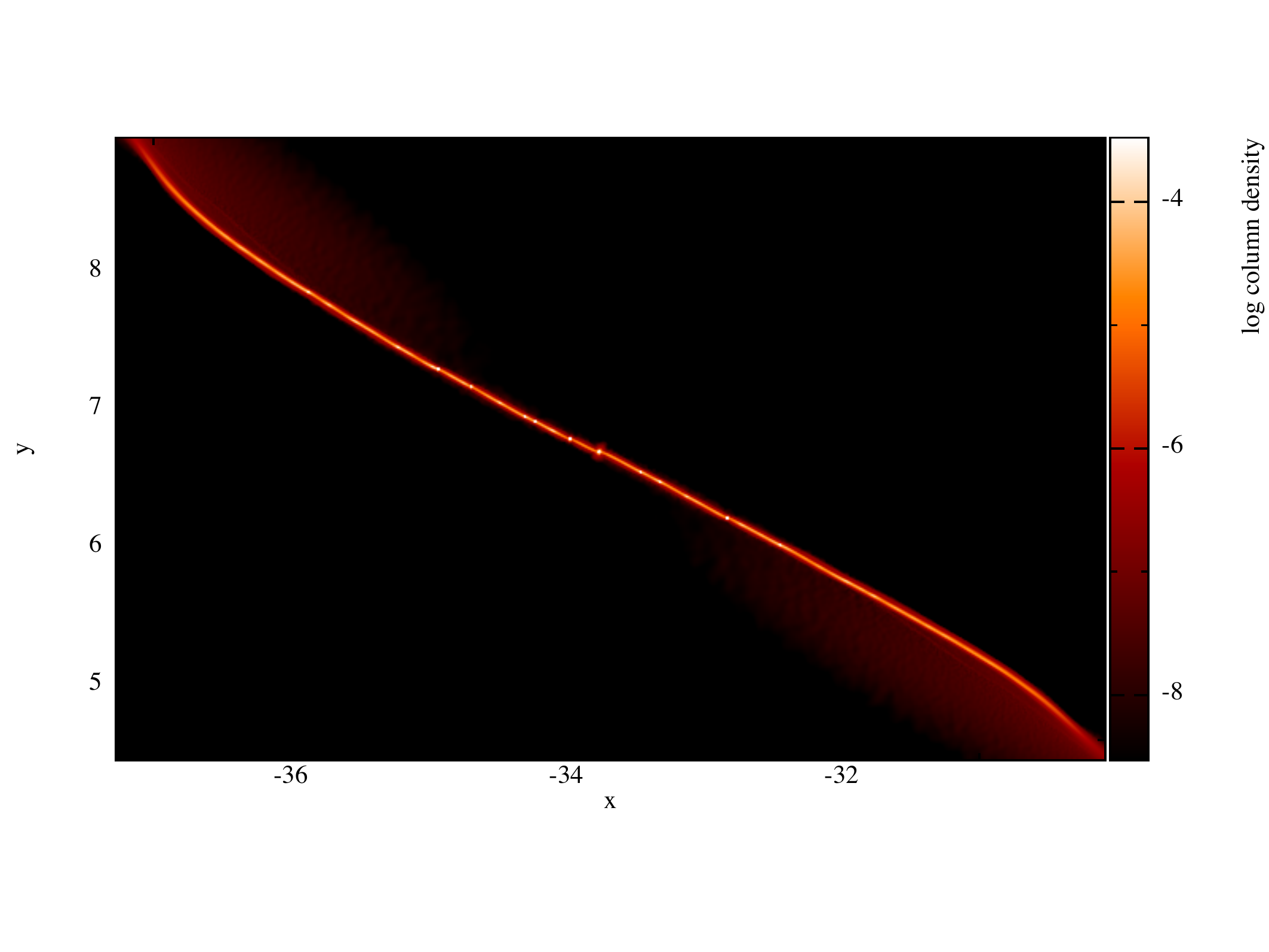}}
\caption{Density map of the stellar debris of Models S3 and S4 
at $t=100$ in the x-y plane over the twelve orders of magnitude 
logarithmic scale in computational unit. Each axis is normalized by the 
tidal disruption radius. The black hole is located at the origin.}
\label{fig:dens}
\end{figure}

Panel (b) of Figure~\ref{fig:raddis} indicates that no fragmentation is apparent 
at $t=100$ for the $n=1.5$ cases, although some structure is seen around the 
highest density region of Model S2. It is apparent from panel (d) that the stellar 
debris remarkably fragments in Model S4, whereas it shows no clear fragmentation 
in Model S3. This demonstrates that a remarkable fragmentation occurs at 
$t=10^5\,{\rm s}\sim1\,{\rm day}$, soon after tidal disruption of the star modeled 
by a $n=3$ polytrope. In panel (d), the dashed line denotes $\rho_{\rm crit}=3\times10^{-3}\,{\rm g\,cm^{-3}}$, which we set as a fiducial minimum value for defining a clump. 
The fourth to sixth columns of Table~\ref{tbl:tdes} shows the flag of whether the 
debris fragments at $t=100$, the number of clumps ($N_{\rm cl}$), and the type of 
corresponding simulation, respectively. Figure~\ref{fig:dens} shows the density 
map of the stellar debris of Models S3 and S4 at $t=100$ in the x-y plane over a dynamic 
range of twelve orders of magnitude. Each axis is normalized by the tidal disruption radius. 
The black hole is located at the origin. The figure demonstrates that there are $\sim20$ 
clumps on the debris in Model S4, whereas it is obvious that there is no fragmentation 
in Model 3. These results are consistent with panel (d).

We find from panel (e) that in Model S3 the gas temperature in the bulk of the debris 
filament (i.e. not in the clumps) is substantially higher than in the Model S4 case, and 
that is why the debris filament does not fragment (because the temperature is typically 
higher at a given density) in Model S3.
It suggests that some mechanism should efficiently work to make the stellar debris cools down. 
One of promising mechanisms is radiative cooling. If the debris is fully ionized, 
its optical depth for Thomson scattering is estimated to be,
\begin{eqnarray}
\tau_{p}=n_p\sigma_{\rm T}r\approx2\times10^9
\left(\frac{\rho}{10^{-3}\,{\rm g\,cm^{-3}}}\right)
\left(\frac{r}{r_{\rm t}}\right),
\end{eqnarray}
where $n_p\approx\rho/m_p$ is the number density of the free electrons, 
$m_p$ is the proton's mass, and $\sigma_{\rm T}$ is the Thomson scattering cross section.  
This implies that radiative cooling is not efficient because the opacity of the gas is so high that 
the photon diffusion time is very long. 
Panel (f) demonstrates that radiative cooling does not work as a debris cooling mechanism almost at all
because the stellar debris clearly fragments in Model S5 as well as Model S4.

\begin{figure}
\centering
\resizebox{\hsize}{!}{
\includegraphics{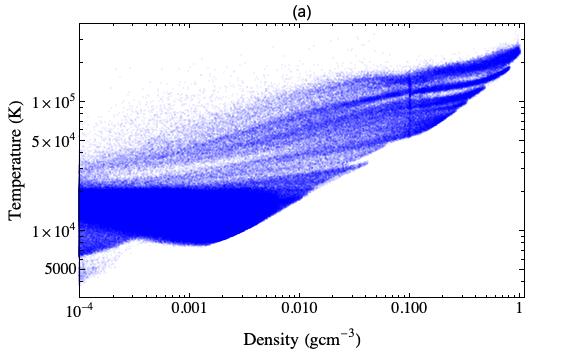}
\includegraphics{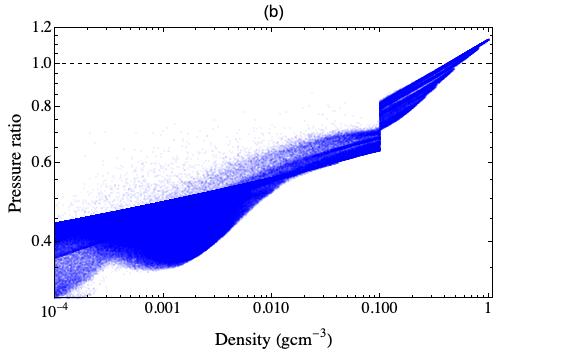}
}\\
\resizebox{\hsize}{!}{
\includegraphics{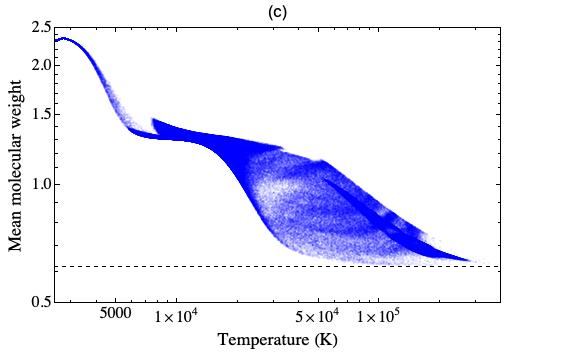}
\includegraphics{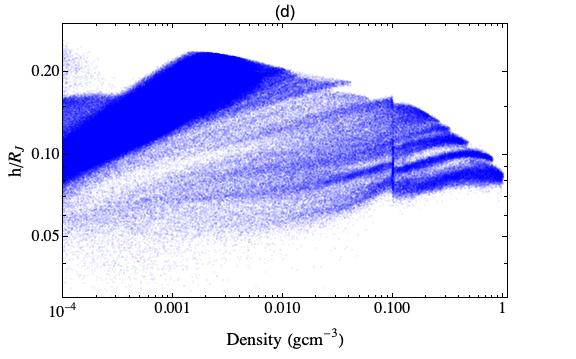}
}\\
\resizebox{\hsize}{!}{
\includegraphics{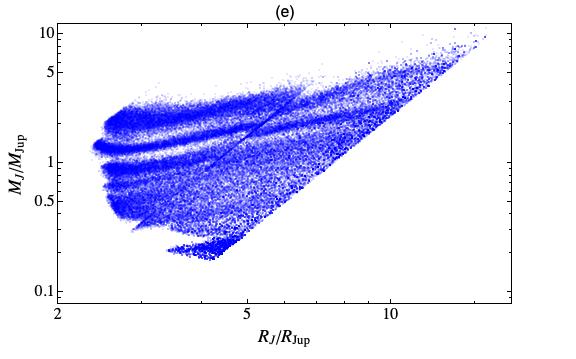}
\includegraphics{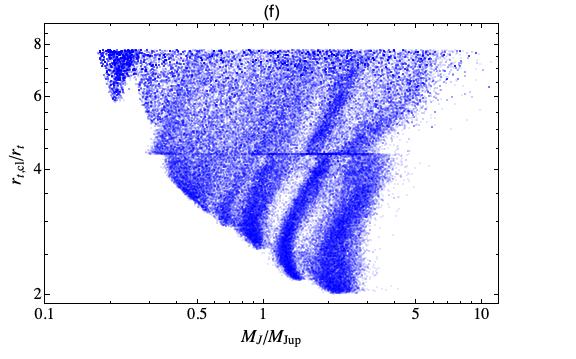}
}
\caption{SPH particle distributions in some phase-spaces of Model 4 at t=100.
In all the panels, each axis is shown on a logarithmic scale. 
(a) Gas temperature-density distribution 
(b) Density dependence of the ratio of 
the pressure computed by equations~\ref{eq:eqs} and \ref{eq:cv2} 
to the pressure of the corresponding ideal gas.
(c) Mean molecular weight-temperature distribution. 
The dashed line denotes $\mu_0=0.62$, which corresponds to 
the mean molecular weight for the fully ionized gas.
(d) Dependence of the smoothing length normalized by $R_{\rm J}$ on the gas density. 
(e) Jeans mass-radius distribution for $\rho\ge\rho_{\rm crit}$, where $\rho_{\rm crit}=3\times10^{-3}\,{\rm gcm^{-3}}$. The Jeans radius and mass are normalized by Jupiter's radius $R_{\rm Jup}$ and 
mass $M_{\rm Jup}$, respectively. 
(f) Dependence of the ratio of stellar to clump's tidal disruption radii (see equation \ref{eq:rtclrt}) on Jeans mass.
}
\label{fig:jeans}
\end{figure}

Figure~\ref{fig:jeans} includes six panels in Model 4: 
panel (a) provides the gas temperature distribution over the density. 
The temperature ranges from $\sim3\times10^3$\,K to $4\times10^4$\,K over $1.0\times10^{-4}\,{\rm gcm^{-3}}\le\rho\lesssim1.0$. The gas temperature tends to increase with the density
Panel (b) depicts the temperature dependence of the ratio of the pressure computed by equation~\ref{eq:eqs} to the pressure of the corresponding ideal gas. The panel indicates that the gas pressure can be significantly weaker than the ideal gas case at some temperature range, where the gravitational collapse can occur. We find from panel (c) that the mean molecular weight changes in two stages: first, it changes due to the dissociation of $H_2$ from $2.38$ to $1.3\sim1.5$ in the range of $3\sim5\times10^3$ K. Next, due to the hydrogen ionization, the mean molecular weight starts at $\sim3\times10^4$ K decreasing to $0.61$, where the gas is fully ionized. In panel (d), the smoothing length is $0.25$ times smaller than $R_{\rm J}$ for more than $10^{-4}\,{\rm g\,cm^{-3}}$. We denote from the panel that the clumps on the debris seen in Figure~\ref{fig:dens} are numerically sufficiently resolved.

In panel (e), the Jeans radius is defined by 
\begin{eqnarray}
R_{\rm J}=\sqrt{\frac{3}{8\pi}}
\left(\frac{R_{\rm g}}{G\mu}\right)^{1/2} 
\left(\frac{T}{\rho}\right)^{1/2},
\label{eq:rj} 
\end{eqnarray}
where we assume that the mean density of a clump 
corresponds to the density of each SPH particle.
With $R_{\rm J}$ estimated from equation~(\ref{eq:rj}), 
the Jeans mass is given by
\begin{eqnarray}
M_{\rm J}=\frac{1}{2}\frac{R_{\rm J}c_{\rm s}^2}{G}=
\frac{1}{2}
\left(\frac{R_{\rm g}}{\mu{G}}\right)TR_{\rm J}.
\end{eqnarray}
From the panel, 
the Jeans mass is distributed over $0.1 < M_{\rm J} / M_{\rm Jup} < 12$ 
and the Jeans radius is distributed over $2 < R_{\rm J}/R_{\rm Jup} < 20$. 
According to \cite{2007ApJ...661..502B}, the observed 
mass-radius relation of giant planets occupies the narrow 
range of  $0.76\lesssim{r_{\rm cl}}/R_{\rm Jup}\lesssim1.4$ 
for $0.35\lesssim{m}_{\rm cl}/M_{\rm Jup}\lesssim1.35$, 
where $r_{\rm cl}$ and $m_{\rm cl}$ are 
the radius and mass of the formed clump, respectively.
For brown dwarfs at age of $\sim{\rm Gyr}$, the observed mass-radius relation 
shows $0.9<r_{\rm cl}/R_{\rm Jup}<1.1$ for $2\le{m}_{\rm cl}/M_{\rm Jup}\le20$ \citep{2011ApJ...736...47B}.
In our simulations, the returning clumps having the orbital period less than $\sim100\,{\rm yr}$ 
are too hot to cool down to be the observed giant planes, while the unbound clumps 
can be potentially freely-floating planets and brown dwarfs if they would keep surviving.

The ratio of stellar to clump's tidal disruption radii is given by 
\begin{eqnarray}
\frac{r_{\rm t,cl}}{r_{\rm t}}=\left(\frac{m_*}{m_{\rm cl}}\right)^{1/3}\frac{r_{\rm cl}}{r_*}
\sim1.02\,\left(\frac{m_*}{M_\odot}\right)^{1/3}
\left(\frac{r_*}{R_\odot}\right)^{-1}
\left(\frac{m_{\rm cl}}{M_{\rm Jup}}\right)^{-1/3}
\left(
\frac{r_{\rm cl}}{R_{\rm Jup}}
\right),
\label{eq:rtclrt}
\end{eqnarray}
where $r_{\rm cl}$ and $m_{\rm cl}$ are the clumpy radius and mass, 
and the tidal disruption radius of a clump is given by
\begin{eqnarray}
r_{\rm t,cl}
=\left(\frac{M_{\rm bh}}{m_{\rm cl}}\right)^{1/3}r_{\rm cl}.
\end{eqnarray}
Panel (f) shows that the tidal radius of the returning clumps 
is larger than the tidal radius of the original star. This suggests 
that all the retuning clumps would be tidally disrupted. 
If the returning clump is disrupted, then the subsequent flare 
would be triggered after the orbital period of the returning debris on the most tightly bound orbit: 
\begin{equation}
t_{\rm mtb,cl}=\frac{\pi}{\sqrt{2}}\left(\frac{M_{\rm bh}}{m_{\rm cl}}\right)^{1/2}
\frac{1}{\Omega_{\rm cl}}
\approx1.2\times10^{8}\,{\rm s}\,
\left(\frac{M_{\rm bh}}{10^6\,M_{\odot}}\right)^{1/2}
\left(\frac{m_{\rm cl}}{M_{\rm Jup}}\right)^{-1}\left(\frac{r_{\rm cl}}{R_{\rm Jup}}\right)^{3/2},
\label{eq:mtb}
\end{equation}
where $\Omega_{\rm cl}=\sqrt{Gm_{\rm cl}/r_{\rm cl}^{3}}$ is the dynamical angular frequency of the clump.
The peak fallback rate is then estimated to be
\begin{eqnarray}
\dot{M}_{\rm cl,max}=\frac{1}{3}\frac{m_{\rm cl}}{t_{\rm mtb,cl}}
\sim8.5\times10^{-5}\,{\rm M_\odot\,yr^{-1}}
\left(\frac{M_{\rm bh}}{10^6\,M_{\odot}}\right)^{-1/2}
\left(\frac{m_{\rm cl}}{M_{\rm Jup}}\right)^{2}\left(\frac{r_{\rm cl}}{R_{\rm Jup}}\right)^{-3/2}.
\label{eq:fbr}
\end{eqnarray}
By taking account of the range of simulated mass and radius of the formed clump, 
which is obtained from panel (b), equation~(\ref{eq:fbr}) implies that the tidal 
disruption of the fallback clumps would be smaller than the Eddington accretion rate $\dot{M}_{\rm Edd}=L_{\rm Edd}/c^2\sim2.2\times10^{-3}\,{M_\odot\,{\rm yr^{-1}}}(M_{\rm bh}/10^6\,M_\odot)$, where $L_{\rm Edd}=4\pi{GM_{\rm bh}}m_{p}c/\sigma_{\rm T}$ is the Eddington luminosity.

\begin{figure}
\resizebox{\hsize}{!}{\includegraphics*[width=10cm]{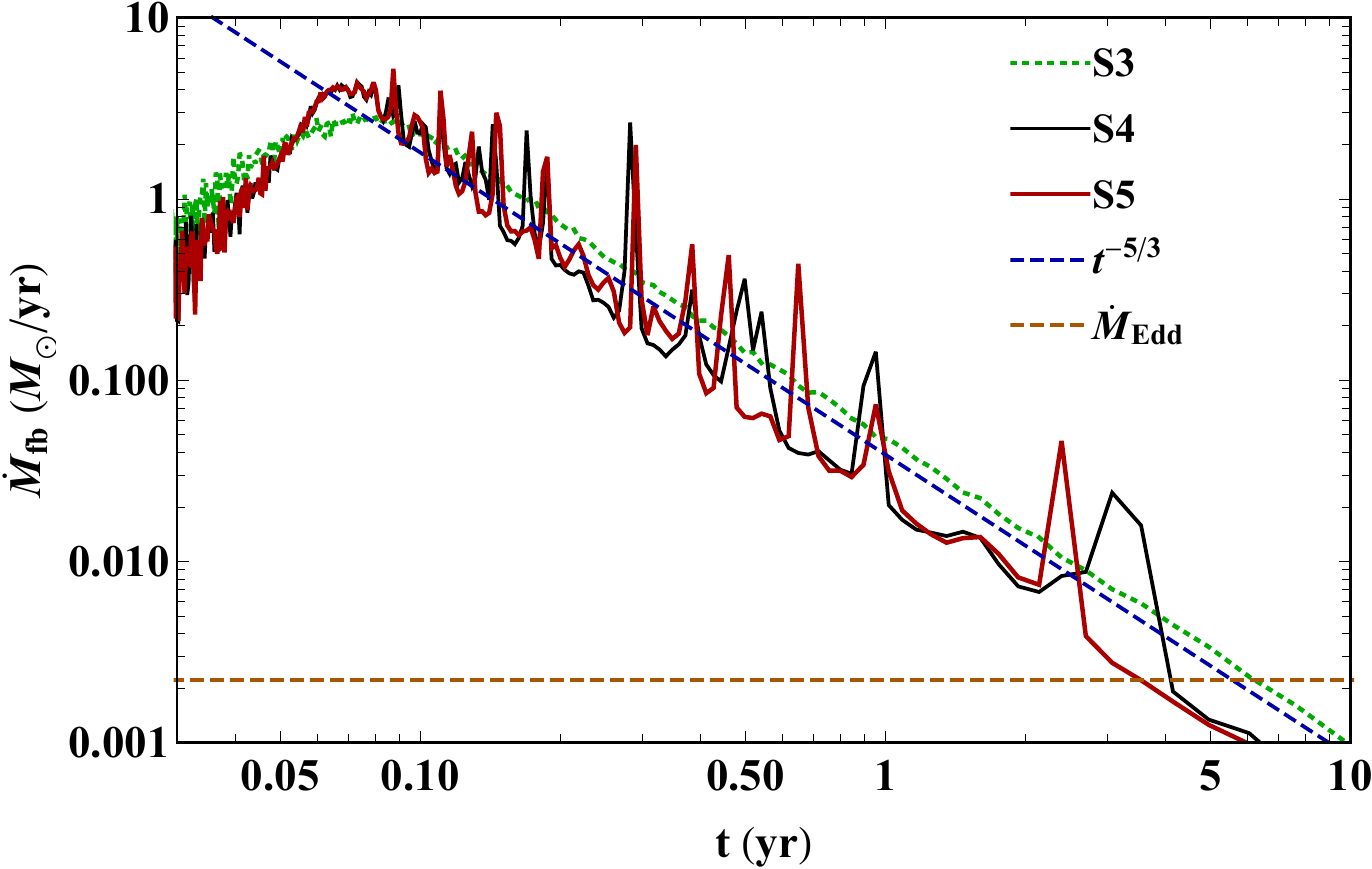}}
\caption{Simulated mass fallback rates of the stellar debris. 
The solid black and red lines show that of Models S4 and S5, 
while the dotted green and dashed blue line represents that of 
Model S3 and the standard $t^{-5/3}$ decay rate.  
The horizontal dashed orange line shows the corresponding 
Eddington accretion rate.
}
\label{fig:fbrate}
\end{figure}

Figure~\ref{fig:fbrate} is the simulated mass fallback rate of the stellar debris 
for each model. The mass fallback rate is defined by $\dot{M}_{\rm fb}=(dM/d\epsilon)(d\epsilon/dt)$, where $\epsilon$ 
is the specific energy of the stellar debris, $d\epsilon/dt\propto{t^{-5/3}}$ because of 
Kepler's third law, and $dM/d\epsilon$ is the simulated differential mass distribution.
The solid black and red lines show the mass fallback rates of Models S4 and S5, 
while the dotted green and dashed blue line represents that of 
Model S3 and the standard $t^{-5/3}$ decay rate. 
The horizontal dashed orange line shows the Eddington accretion rate. 
The mass fallback rates of Models S4 and S5 decays with several pronounced spikes 
from the early to late time. The spikes in the light curve 
can be used as evidence for the clump formation in TDEs.

%
\section{Discussions and Conclusions}
\label{sec:con}
%
%
We have performed the RHD simulations of tidal disruption of a star 
on a parabolic orbit by a SMBH by using a 3D SPH code with radiative transfer. 
Our conclusions are as follows:
\begin{enumerate}
\item Tidally disrupted stars fragment and 
form $\sim20$ clumps soon after tidal disruption ($\sim1\,{\rm day}$ for an SMBH of $10^6\,M_\odot$). 
This formation is triggered by the endothermic processes through ionization and dissociation of helium and hydrogen, decreasing the gas pressure compared with the adiabatic gas case, leading to local gravitational collapse. Radiative cooling is ineffective almost at all in causing fragmentation because 
the stellar debris is highly optically thick.
\item
The fragmentation does not occur in tidal disruption of a star modeled as a $n=1.5$ polytrope.
This is because the highest density is an order of magnitude lower than that of $n=3$ case.
%
%
\item The mass fallback rate decays with time but shows several pronounced spikes 
due to the formation of clumps. The detection of such spikes can serve as a smoking-gun 
for clump formation in stellar debris.
\item The clumps on unbound debris can be potentially freely-floating planets and brown dwarfs.
\end{enumerate}
%
%
%
%
We have also tested the dependence of 
the debris fragmentation for the $n=3$ case on the penetration factor $\beta$, 
and found that it occurs only for $\beta\gtrsim2.2$. This condition resembles the condition 
for avoiding the partial tidal disruption, where the core of the star survives 
\citep{2017A&A...600A.124M}. 
We find that if the fraction of SPH particles making of the central, biggest clump 
is less than one-third of the total particle number, the debris can fragment. 
The degree of central concentration determines whether the stellar 
debris fragments or not. This is possibly due to the suppression of the self-gravitating instability 
by the central clump. The suppression condition is given by the balance 
between the self-gravity of the smaller clump and the gravity acting on it 
from the central clump as $Gm_{\rm cl}^2/r_{\rm cl}^2\,\lesssim\,{G}m_{\rm c}m_{\rm cl}/r^2(H/r)$, 
where $m_{\rm c}$, $H$, and $r$ is the mass of the central clump, the debris scale hight, and 
the distance between the the central and the other clumps. 
Because this results in $r/r_{\rm cl}\lesssim(m_{\rm c}/m_{\rm cl})^{1/2}(H/r)^{1/2}\sim\mathcal{O}(1)$, 
we find the suppression by the central clump cannot work. 
If the originally approaching star is more massive than 10$R_\odot$, then each 
clump's mass should correspond to a star. 
This is a promising mechanism for making low-mass stars around a SMBH or an IMBH. 


\newpage

%
%

%
%

%
%

%
%

%
%

\acknowledgments
This research has been supported by Basic Science Research 
Program through the National Research Foundation of Korea (NRF) 
funded by the Ministry of Education (NRF-2016R1A5A1013277 
and 2017R1D1A1B03028580 to KH), by the European Research 
Council under the European Community’s Seventh Framework 
Programme (FP7/2007- 2013 grant agreement no. 339248 MRB), 
and by the Black Hole Initiative at Harvard university which is 
funded by JTF and GBMF (AL). The numerical simulations were 
performed and also supported by the National Supercomputing 
Center with supercomputing resources including technical support 
(KSC-2019-CRE-0082 to KH).

\end{document}